# Quality assessment of in situ plasma etched diamond surfaces for CVD overgrowth

J. Langer, V. Cimalla, M. Prescher, J. Ligl, B. Tegetmeyer, C. Schreyvogel and O. Ambacher

*Fraunhofer Institute for Applied Solid State Physics (IAF), Tullastraße 72, 79108 Freiburg, Germany*

**Abstract**

In situ plasma etching is a common method to prepare diamond substrates for epitaxial overgrowth to effectuate higher quality. However, there is no practical, direct, qualitative method established so far to assess the performance of the etching pretreatment. We propose an optimization of the pretreatment process on grounds of high-resolution X-ray diffraction measurements to judge the structural quality gain of the diamond substrates and the effectiveness of the polishing-induced subsurface damage removal. The obtained data shows, that parameters like thickness nor misorientation angle of the diamond substrates seem to influence the gain of the structural quality. The process duration, however, is an important key factor, when the amount of material removal and the arising roughness are discussed. Furthermore, the impact of the oxygen-to-hydrogen ratio is examined. And with rising oxygen percentage, the structural quality gain remains similar, only the overall as well as local mean roughness increases exponentially. Within the utilized reactor setup, the best results are obtained after a 20-minute in situ hydrogen plasma-etching step. The optimal pretreatment process, however, will always embody a tradeoff and needs to be optimized for each reactor type. Due to the introduced method a better evaluation and comparison of the achievements is accomplishable.

*Keywords:* synthetic diamond, pretreatment process, in situ etching, structural analysis, chemical vapor deposition

## 1. Introduction

For the realization of diamond-based optical as well as electronic devices, reproducible and affordable high quality standards are demanded. Homoepitaxial microwave plasma enhanced chemical vapor deposition (MW-PECVD) is currently the most adequate method to prepare tailor-made diamond with precisely defined composition, purity and structural quality. To meet the required demands, the reactor setup as well as the growth conditions need to be optimized. Yet another important criterion within this production chain is the starting material itself. The dislocations within the bulk crystal can thread through the growth layer and surface defects can additionally cause nucleation of dislocations [1, 2, 4]. In this connection, it is necessary to ensure the substrate quality of the diamond bulk and its surface finish. Over the past two decades, different kinds of diamond material suitable for CVD overgrowth were investigated on their defect densities and structural quality [2, 3, 4]. Pristine diamond substrates with a very low concentration of undesired contaminants and a very high structural quality are, however, rather rare and only accessible to a certain extent. The issue in general hereby is the lack of defined quality standards on the commercially available diamond substrates, which are fluctuating largely in the structural quality as well as the contaminant concentrations. Polishing is a common method to provide at least a reproducible surface for epitaxial overgrowth. It induces, however, additional subsurface damage, which can be controlled by an optimized polishing process with low damage infliction [5] and can further be reduced with an optimized pretreatment etching process of the substrates. Over the years of research, different methods for the pretreatment of diamond substrates were proposed. Methods like reactive ion etching (RIE), electron cyclotron resonance (ECR) or inductively coupled plasma (ICP) etching were investigated and major improvements for the CVD overgrowth were demonstrated [6, 7, 8, 9]. An in situ pretreatment process, however, features several advantages over the previously mentioned methods like a reduced expenditure of time, energy and cost as well as the prevention of an exposure to air and therefore contaminants during transfer. Thus, literature states an in situ plasma-etching step as an essential pretreatment process prior to a high quality CVD overgrowth [10, 11]. Unfortunately, only a few results of systematic approaches to better understand the influence of different parameters on the pretreatment process are published so far: Prevalent findings are that the etching rate is exponentially increasing with rising substrate temperature and an additional incline in misorientation angle is as well leading to an increase of the etching rate [12, 13]. The importance hereby is the actual evaluation of the effectiveness of the polishing-induced subsurface damage removal. Usually, it is indirectly performed by subsequent epitaxial overgrowth [6, 7, 8, 14]; however, this method implicates various collateral factors of influence, which cannot be accounted for. Therefore, a direct method to assess the performance of the etching pretreatment would lead to a significant improvement and simplification on its optimization. Besides in situ low-coherence interferometry [13], the only applied direct evaluation approach on in situ plasma etching of diamond substrates is to determine the saturation of the etch pit density [3], which, however, implies using a method that produces a very rough surface. Thus, this work proposes Omega scans by high-resolution X-ray diffraction as nondestructive tool to effectively optimize the in situ plasma etching pretreatment process of diamond substrates on grounds of structural quality gain. This technique is already an established method to evaluate the polishing-induced subsurface damage removal in fields of research on other material systems like silicon carbide [15], silicon [16] as well as aluminum nitride [17].

For the process optimization, the influence of different parameters like the process duration and the oxygen-to-hydrogen ratio of the experimental runs as well as the thickness and the misorientation angle of the samples are investigated. These parameters are varied over a pertinent range to see if and how they influence the resulting quality of the diamond substrate after etching. The quality assessment is further elaborated by an optical inspection and white-light interferometry measurements.

## 2. Experimental Setup

The pretreatment process was optimized by using (001) HPHT and CVD diamond substrates. These substrates were carefully pre-characterized by high-resolution X-ray diffraction (HRXRD) Omega scan measurements of the 004 reflection. A PANalytical X'Pert Pro MRD system equipped with Cu radiation ($\lambda$ = 154.06 pm) was used for these measurements. On the primary beam side, a multilayer mirror and a double bounce Ge (220) monochromator was inserted into the beam path, while for the diffracted beam side a triple bounce Ge (220) analyzer was installed. The mean roughness (Ra) of the samples after polishing was determined via white-light interferometry to be mainly well below 3 nm Ra (WYKO NT1100). Additionally, optical microscopy was conducted with and without crossed polarizers (Leica DMRM). Afterwards, the substrates were sorted by their thickness and misorientation angle to be assembled for different experimental series regarding process duration, thickness variation, misorientation angle variation and variation of the oxygen content in the total gas flow.

Prior to etching, all substrates underwent an extended cleaning process, which included boiling in nitro sulfuric acid and several steps of organic solvent cleaning. It is of great importance that the substrates are spotless, because any residue of metal or other contaminants can influence the plasma interaction with the substrate and can cause additional unpredictable quantities of local etching.

The plasma etching experiments were performed in an ellipsoidal shaped CVD reactor with a 2.45 GHz microwave frequency and equipped with a 6 kW microwave generator [18].

The post-characterization was executed with the same measurement setups and parameters to receive comparable results from before and after the plasma etching pretreatment process. The etching rate was calculated from the weight difference of the sample measured by an ultra-microbalance (Mettler Toledo UMX5).

## 3. Results & Discussion

*Hydrogen plasma etching pretreatment*

The process duration of the plasma etching step is an important aspect of an optimization process. Several successive etching steps were performed on one HPHT substrate to obtain data on the etching rate and the evolution of the quality over the time of etching.

The effective etching rate monotonically decreases from about 1 µm h$^{-1}$ after 5 min of etching to less than 100 nm h$^{-1}$ after 40 min. It should be noted that these measurement results are subject to large tolerances due to the limitation of thickness determination by weighing. Nevertheless, they show a clear trend and are in a good qualitative agreement to the observation of increased initial etching of a "defected layer" by Yurov et al. 2017 [13]. It implies that a surface-near region of the diamond has a substantially high defect density, which amplifies the etching in hydrogen. The equilibrium-etching rate of the bulk diamond after longer etching times is below 100 nm h$^{-1}$ in agreement with literature [13].

The removal of the "defected layer" was further investigated by high-resolution X-ray diffraction measurements. In Figure 1 a), the black Omega scan of the 004 reflection displays the quality of the (001) substrate as received. The sample was pretreated by hydrogen plasma etching in the following time intervals: 5 min, + 5 min, + 10 min, + 20 min, + 560 min; adding up to 600 min of etching time in total. The well addressed FWHM of the Omega scan measurements did not change substantially: The substrate exhibited a FWHM of 11.0" and was slightly reduced to 10.1" after 40 min of hydrogen etching and marginally increased to 11.1" again with the final etching step (Figure 1 b). However, observing the diffuse scattering at the slopes around the 004 reflection of diamond provides additional information on the crystal lattice distortion [19]. For a rough quantification of the diffuse scattering, the full width at ten thousandth of the maximum (FWTTM) was analyzed. The data is displayed in Figure 1 c), which shows a logistic decline, allowing to deduce a significant reduction of the subsurface damage. The results illustrate that the longer the sample is etched the steeper the slopes get and therefore the quality of the sample increases. Furthermore, the mean roughness (Ra) of the sample increases exponentially with longer etching duration. Therefore, when considering both the structural quality gain and the mean roughness of the diamond sample, a reasonable compromise lies in between. Within our reactor setup, an optimal tradeoff for the pretreatment process is reached after 20 min of in situ hydrogen plasma etching, where the measurable structural quality gain begins to saturate while the surface roughness only doubles.

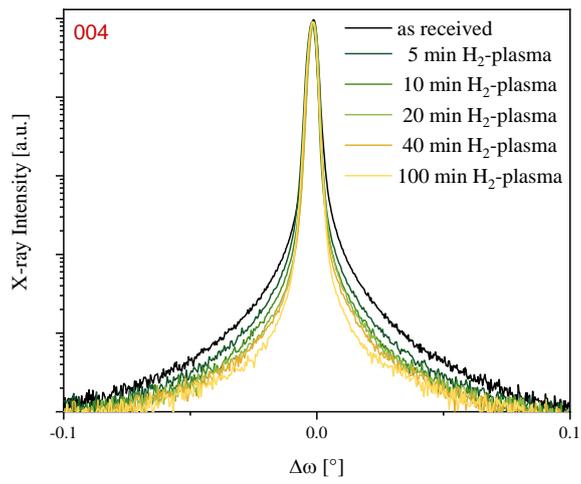

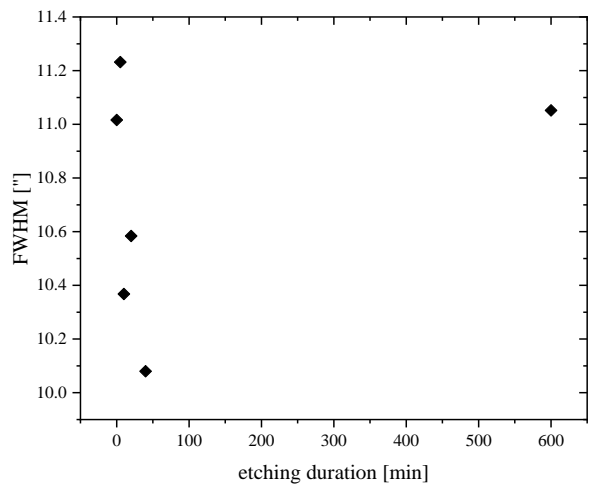

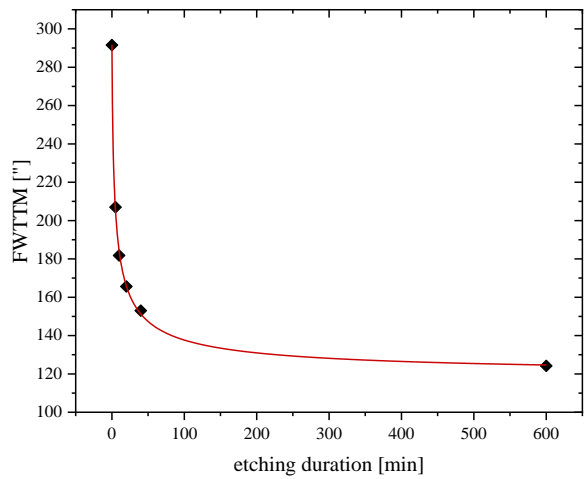

Figure 1. a) Omega scan data of a successive etched sample. b) FWHM as a function of etching duration. c) FWTTM as a function of etching duration.

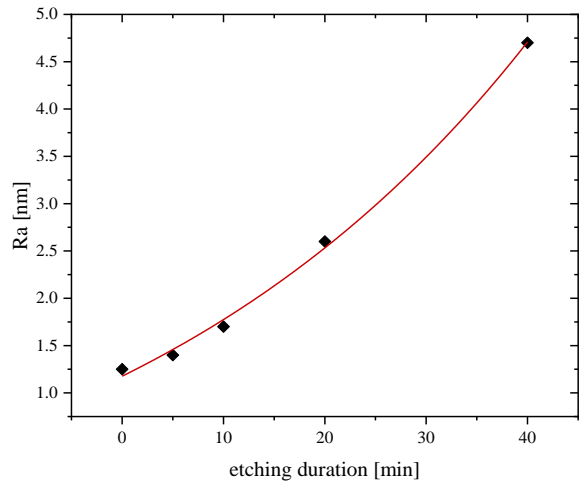

Figure 2. Evolution of the mean roughness (Ra) of the plasma etched substrate as a function of etching duration.

Continuing with the optimized duration of the etching process, the influence of the thickness, how far the substrate reaches into the plasma, as well as the misorientation angle of the HPHT diamond substrate on the quality gain after pretreatment were investigated. The results on the predominance of the effect from these parameters should enable an estimation on accuracy for further etching experiments. The two experimental series were conducted on diamond substrates with rather poor quality of the surface finishing ensuring the sensitivity of the method when evaluating the arising tendencies. However, no correlation between the thickness of the sample and the etching rate nor the quality gain could be substantiated. The same findings apply for the data of the misorientation angle series. It should be noted that Ivanov et al [12] found a strong dependency of the etching rate on the misorientation angle for bulk diamond, which was substantially lower than the etching rate of the "defected layer". Thus, the high density of the polishing-induced subsurface defects was dominating the etching of the diamond substrates and obscured the actual impact of the misorientation angle. The only real determinable trend was the correlation between the structural quality given by the FWHM of the Omega scan measurements before and after the etching process. The data is displayed in Figure 3. The maximum quality gain of the pretreated substrates calculated by a linear fit through the data averages out at around 35 %.

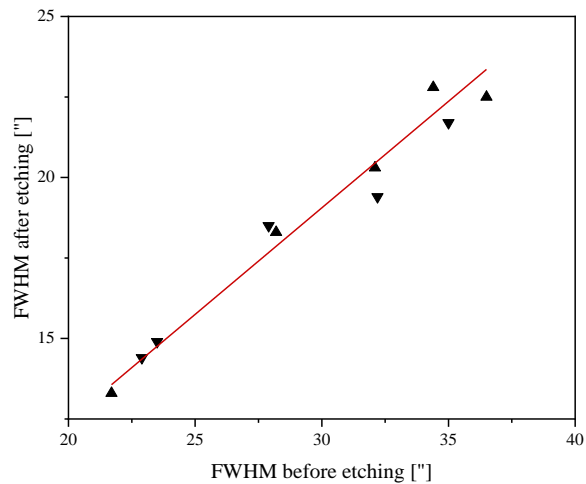

Figure 3. FWHM of Omega scans from hydrogen plasma etching experiments: thickness series (triangle up) & misorientation angle series (triangle down).

*Oxygen-hydrogen plasma etching pretreatment*

Literature depicts that the quality gain through a decent plasma pretreatment process can be even further enhanced by adding oxygen as additional etchant to the plasma [3, 6, 14, 20]. To prove this working hypothesis, a series with varying oxygen-to-hydrogen ratios was conducted: starting out from 0 % up to 1.5 % oxygen in hydrogen. These pretreatment experiments with oxygen were conducted on a batch of high quality CVD substrates with FWHM below 8.5" to not distort the outcome on grounds of the findings from the data displayed in Figure 3. The FWHM of the Omega scans did not change, slight variation of minimal FWHM values are solely due to the resolution limits of the diffractometer. However, when correlating the Omega scan data from before and after the etching experiment, a peak sharpening due to a reduction of diffuse scattering can still be observed despite the already high structural quality of the diamond substrates depicted by the very low FWHM (Figure 4). This tendency can be noted in all the results throughout the whole oxygen-hydrogen etching series. Ultimately, the quality gain seems to be consistent while independent of the oxygen percentage used in the pretreatment process.

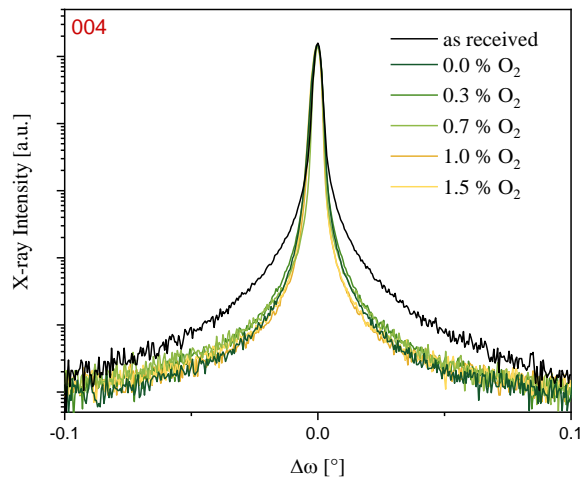

Figure 4. Omega scan data of an untreated CVD diamond substrate (black) in comparison to pretreated ones, which are plasma etched with different oxygen-to-hydrogen ratios (green to yellow).

The difference in the etching experiments series with increasing oxygen-to-hydrogen ratio becomes only visible during optical characterization and the determination of the etching rate. The latter is contrasted with the rising oxygen percentage of the different pretreatment processes in Figure 5. The data illustrates a strong increase of the etching rate with higher oxygen-to-hydrogen ratios. Figure 6 a) - d) display microscope images with differential interference contrast (DIC) of four plasma-pretreated substrates. From these images, it is evident that the influence of oxygen rises with its percentage: The higher the oxygen concentration the more etch pits appear. The mean roughness (Ra) of the in situ oxygen-hydrogen plasma etched pretreated substrates are shown in Figure 7, supporting the impression of an increasing roughness with higher oxygen concentrations from the DIC microscopy images shown in Figure 6 a) – d). However, not only the overall roughness is effected based on the rising amount of etch pits, but also the depth of the single etch pits became greater. With 1.5 % oxygen in the plasma, the resulting etch pits are already up to 13 µm deep. Oxygen as an etchant within the plasma is effecting the surface more locally in contrast to pure hydrogen plasmas. This leads to the finding that oxygen within the plasma pretreatment process may not certainly result in a higher quality gain, but in this case only in a rougher surface.

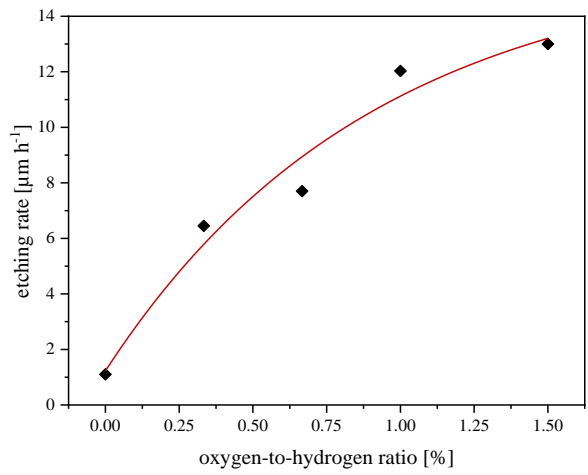

Figure 5. Evolution of the etching rate as a function of the oxygen percentage within the plasma.

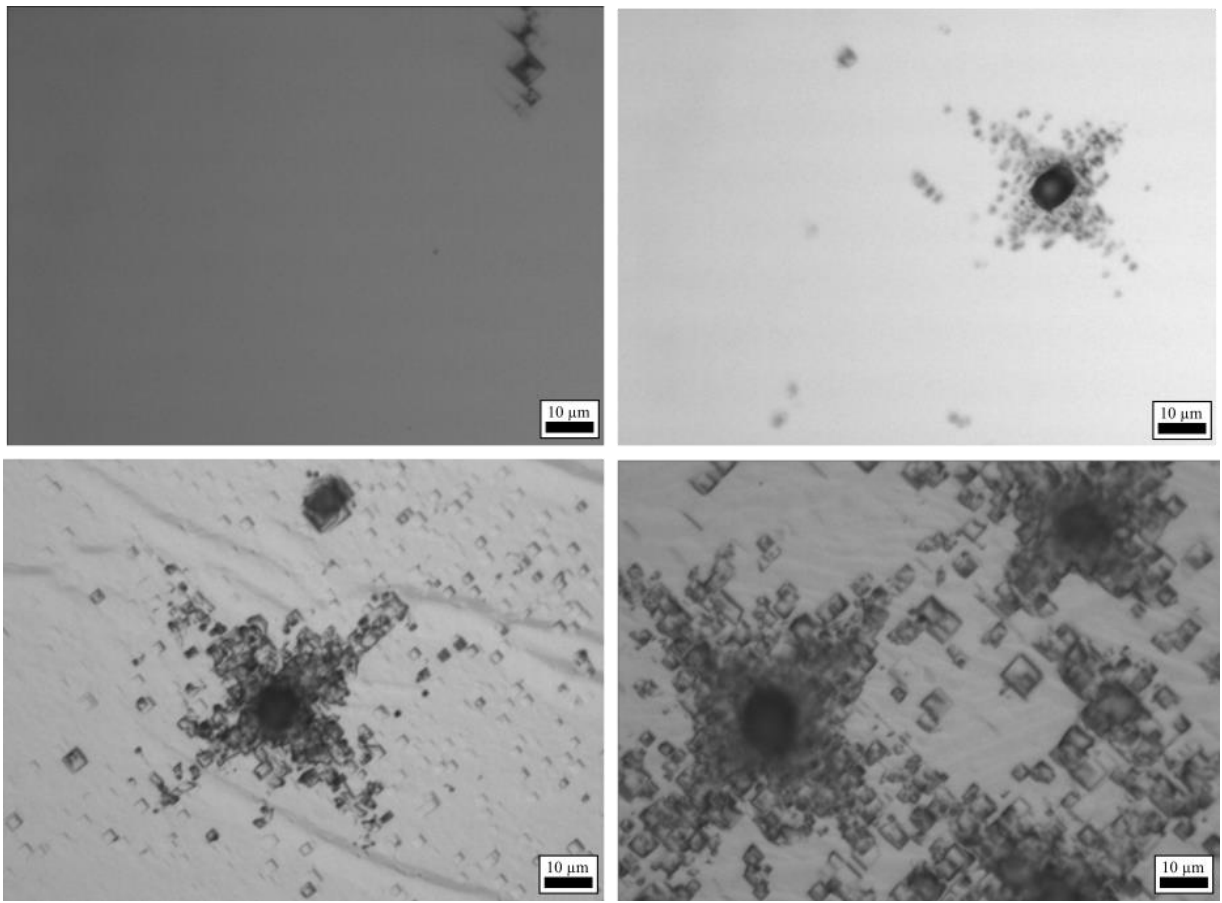

Figure 6. DIC microscopy images of plasma etched diamond substrates with different oxygen-to-hydrogen ratios: a) 0 %, b) 0.33 %, c) 1 % and d) 1.5 %.

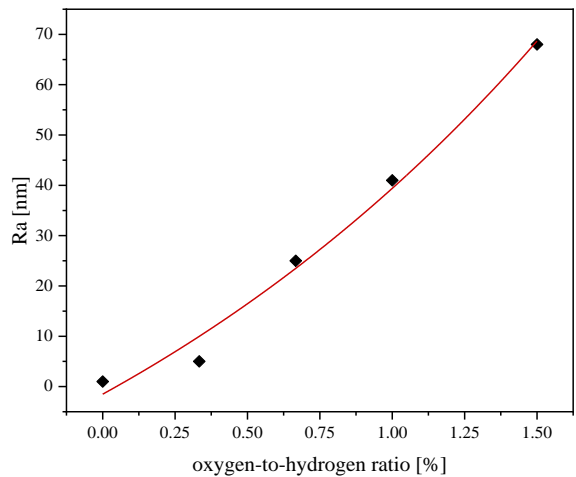

Figure 7. Evolution of the mean roughness (Ra) of the plasma etched substrates as a function of the oxygen percentage in the plasma.

## 4. Conclusion

The etching rate of the diamond substrates with polishing-induced subsurface damage is substantially higher than for bulk diamond indicating a high density of defects. The appearance of such defects can be analyzed through Omega scan measurements by high-resolution X-ray diffraction. Depending on the initial quality of the substrates, an optimization can be achieved with the reduction of the FWHM. Speaking in terms of FWHM improvement a quality gain, up to 35 % can be obtained. Especially for high-quality samples, however, the determination of the FWHM is not informative enough. The displayed data shows that the increasing steepness of the Omega scan slopes, approximately quantifiable by the FWTTM, is additionally very significant. Generally, the FWTTM decreases during etching until it reaches an equilibrium state, which determines the entire removal of the polishing-induced subsurface damage. Within the utilized reactor setup, the best tradeoff between etching duration, mean roughness and structural quality gain was attained after 20 min of etching without the addition of oxygen to the plasma. Oxygen as admixed etchant only increases the etching rate and the overall as well as the local surface roughness without further improving the structural quality of the diamonds. Furthermore, the thickness and the misorientation angle of the diamond substrates only show a negligible impact on the removal of the polishing-induced subsurface damage. A much more dominant factor on the in situ etching pretreatment process is the structural quality of the diamond substrates prior to etching.

The presented data indicates that high-resolution X-ray diffraction is a valid direct qualitative method to determine the reduction of the polishing-induced subsurface damage on diamond substrates by an in situ plasma etching process and offers a possibility to establish a quality assessment criterion. This proposed evaluation method further enables detailed studies about the impact of the subsurface damage on the homoepitaxial growth of high quality diamond.

**Acknowledgments**

The authors acknowledge funding from the German federal ministry for education and research Bundesministerium für Bildung und Forschung (BMBF) under grant number 13XP5063.

**Data availability**

The data that support the findings of this study are available from the corresponding author upon reasonable request.